# Terahertz response of biological tissue placed on a silicon nanostructure

K.B. Taranets[1], M.A. Fomin[1], L.E. Klyachkin[2], A.M. Malyarenko[2], N.T. Bagraev[2], A.L. Chernev[2]
[1]Peter the Great St. Petersburg Polytechnic University St. Petersburg, 195251, Russia
[2]Ioffe Institute, St. Petersburg, 194021 Russia

**Abstract**

Here we present the first findings on the resonance response of DNA oligonucleotides deposited on silicon nanosandwich (SNS) and living bio-tissue to the THz irradiation that allow their identification by measuring the changes of the longitudinal conductance and the lateral voltage within frameworks of the SNS prepared in the Hall geometry. The mechanism of the THz response is discussed, with the model of the Shapiro steps related generation. The THz resonance response from living bio-tissue under the THz radiation is also applied to the definition of oncological diseases. The results obtained form the basis of the express diagnostics in practical medicine.

**Introduction**

The actual task of physics and technology of semiconductors is the search for new methods for detecting and identifying DNA oligonucleotides. New approaches to this problem have allowed to improve the existing biotechnological methods [1] and create a number of new ones [2,3]. Therefore, of particular interest are studies aimed at finding new approaches to the identification of oligonucleotides. It is believed that the concept of personal medicine will be implemented when the cost of a comprehensive analysis of the human genome will fall so that it becomes available at a mass level. It should be noted that most modern methods for analyzing genetic information are based on sequencing of the genome, the possibility of which, in turn, is due to the development of technical methods for detecting an increase in oligonucleotide by one nucleotide. However, it should be noted that sequencing is a nucleotide technology for identifying and analyzing an oligonucleotide, while it is possible to identify the properties of oligonucleotide sequences as a whole. To do this, it is necessary to investigate the properties of oligonucleotide molecules that can characterize the length and primary structure of the oligonucleotide. These properties include the dielectric properties of DNA. Based on a comparison of experimental data on conduction studies, the nucleotide composition and length of the oligonucleotide play a primary role in the formation of the dielectric properties of these biomolecules.

An alternate approach to the problem of sequencing oligonucleotides with a large number of repeated elements is associated with the use of membranes, through the use of which single ion counters are created (according to the Coulter counter scheme [4]), and carrier tunneling currents are analyzed when nucleotides pass through integrated nanopores. It turned out that the change in the ion current can be used to judge the type of nucleotide passing through the nanopore [5–7].

Nevertheless, the fundamental problem of sequencing using nanopores in membranes is limited control of the time of oligonucleotide passage through the nanopore [7].

It should be noted that special attention is paid to the use of THz pumping to develop methods for analyzing DNA oligonucleotide's properties and also for methods of their express diagnostics.

Recent studies have discovered THz spectral properties of various proteins and their compounds, including DNA oligonucleotides. Also characteristics of the terahertz (THz) radiation of healthy biological tissue have been shown to differ from the corresponding indices of its sites affected by oncological diseases [8].

In this regard, the use of compact sources of THz radiation, among which silicon nanosandwiches (SNS) can be distinguished, is relevant. Those nanosandwiches are able to emit in a wide THz range when passing electric current.

Using IR Fourier spectroscopy technique with the Bruker Vertex 70 and the IFS 125HR FTIR spectrometers by measuring the transmission spectra, the electron-vibrational bond modes in protein compounds in the THz frequency range were studied thereby opening up the possibility of using THz sources and recorders in practical medicine and biology that was the goal of this work which is to identify of DNA nucleotides by the terahertz analyzing the electrical characteristics as well as to develop early diagnosis in terms of the reflection of the THz radiation from biological tissue.

**Methods**

The SNS represents an ultra-narrow p-type silicon quantum well (*p*-Si–QW) confined by δ-barriers heavily doped with boron ($5 \times 10^{21}$ cm$^{-3}$) on an *n*-Si (100) surface in which a high carrier mobility is achieved (Figs. 1a–1d) [9, 10]. These *p*-Si–QWs are formed on *n*-Si (100) substrates during preliminary oxidation and subsequent short-term boron diffusion from the gas phase [9, 11-13]. It was shown that boron atoms in the δ-barriers form trigonal dipole centers (B$^+$–B$^-$) due to the negative-U reaction $2B° \rightarrow B^+ + B^-$ [11, 14, 15], whose crystallographically oriented sequences form edge channels responsible for conduction in *p*-Si–QWs [14]. The two-dimensional hole density was determined by measuring the field Hall dependences as $3 \times 10^{13}$ m$^{-2}$ [10, 16, 17].

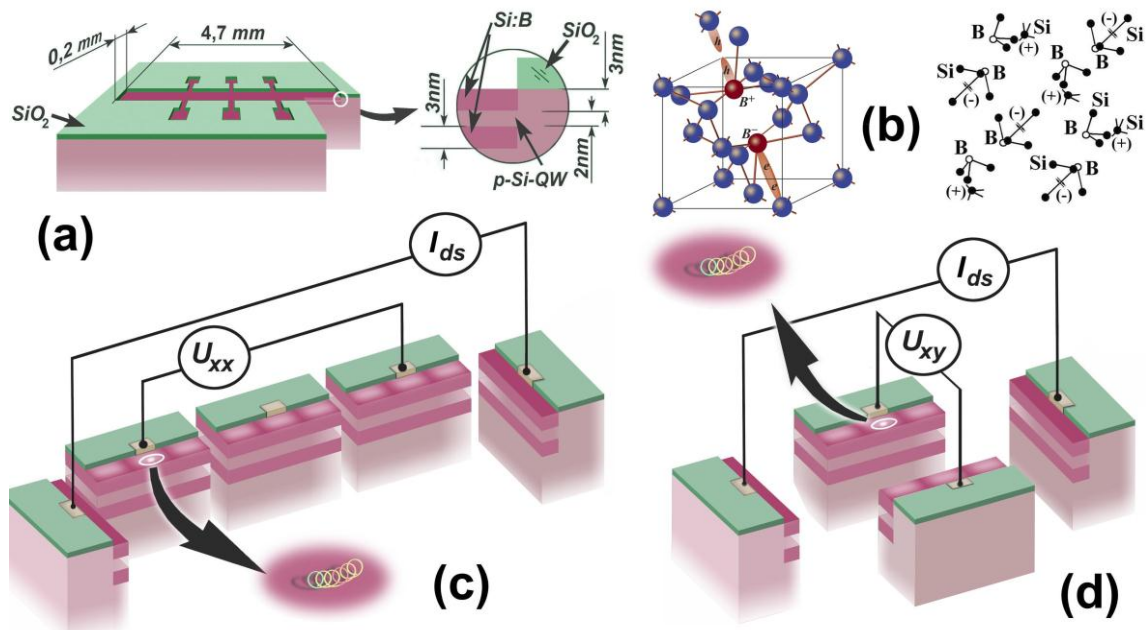

**Fig. 1.** (a) Schematic representation of a SNS with characteristic sizes. (b) Dipole trigonal boron center (B$^+$–B$^-$) with negative correlation energy and chains of dipole boron centers in δ-barriers confining p-Si–QWs. Measurement scheme for studying the oligonucleotide response depending on (c) the conductance and (d) the transverse potential difference on the stabilized source– drain driving current.

Furthermore, it was shown that *p*-Si–QW edge channels under conditions of longitudinal current are efficient sources of THz and GHz (gigahertz) radiation caused by the presence of negative-U dipole boron centers (Figs. 1b–1d) [14]. In this case, the frequency f$_0$ = 9.3 GHz is most pronounced in the centimeter range, since the *p*-Si–QW confined by δ-barriers heavily doped with boron is shaped as a microcavity with a characteristic length corresponding to its value of *d*, where $d = \lambda/2n = 4.72$ mm, *n* is the refractive index (for silicon, $n = 3.4$), f$_0$ = $c/\lambda$ = 9.3 GHz. The presence of 9.3-GHz radiation from the SNS was confirmed via various experimental techniques: this frequency appears in the modulation of IR photo- and electroluminescence spectra measured by IR Fourier spectroscopy [9, 16, 17]. It should be noted that 9.3-GHz radiation generation in the source–drain longitudinal current mode (see Fig. 1a) allows measurements of the spectra of the electrically detectable electron paramagnetic resonance of various point centers in the SNS by measuring the magnetoresistance in the absence of external microcavities, a microwave source and a recorder [14]. In the present study, 9.3-GHz radiation is identified by measuring the change in the SNS conductance under conditions of sweeping the stabilized source– drain current with an extremely small step (10– 100 pA) (Fig. 2).

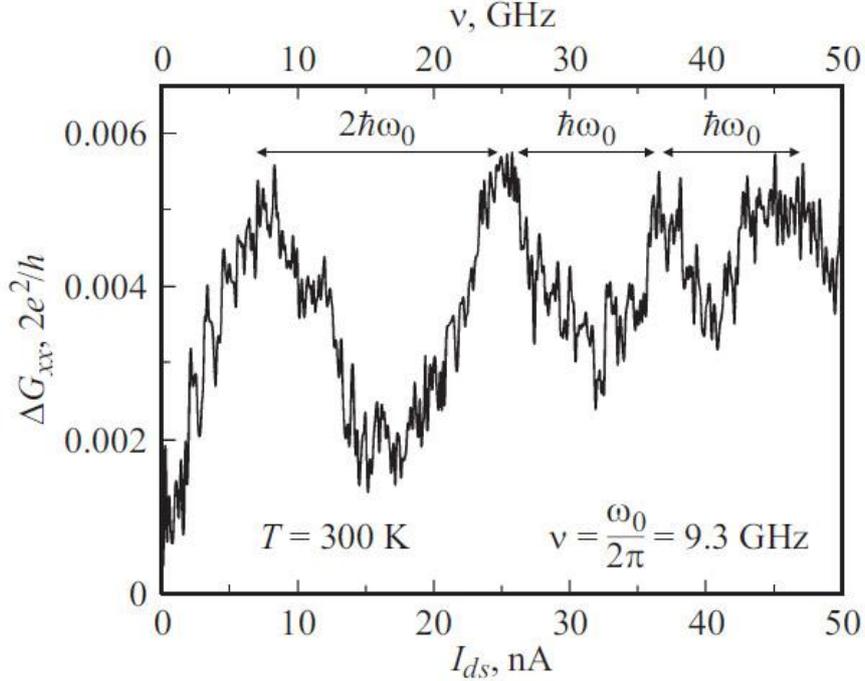

**Fig. 2.** Dependence of the change in the conductance on the stabilized source–drain driving current under conditions of GHz-radiation generation in the SNS.

The appearing characteristic conductance peaks are caused by the depression of electron–electron interaction that is due to the exchange interaction of single holes with dipole centers with negative correlation energy, which leads to the formation of chains of quantum harmonic oscillators containing single exchange-coupled holes within the edge channels [17]. It is these quantum harmonic oscillators in the edge channels that are sources of GHz–THz radiation whose frequency can vary depending on the source–drain current, and the intensity is determined by the mismatch with the intrinsic frequency of the embedded microcavity [14]. Within the aforesaid, to analyze the data shown in Fig. 2, let us determine the minimum source–drain current corresponding to 9.3-GHz radiation generation by such an individual harmonic oscillator, using the uncertainty relation $\Delta E \Delta \tau \geq \hbar$, where $\Delta \tau = e/\Delta I$, $\Delta E = h\nu = \hbar\omega$. From this it follows that $\hbar\omega * e/\Delta I \geq \hbar$ or $2\pi f * e/\Delta I \geq 1$. Thus, the minimum current required for the onset of generation is $\sim 2\pi f * e = 9.3$ nA (Fig. 2). We note that the insufficiently small step in the source–drain current variation can introduce a discrepancy into the ratio of amplitudes of peaks caused by close harmonics (Fig. 2). As the source–drain current increases, the suppression of peaks corresponding to high-frequency radiation is observed (Fig. 2). However, this effect is leveled by the inclusion of an ever increasing number of harmonic oscillators ($n$ pieces in the general case) to the generation process. In the case where the electron–electron interaction is sufficiently suppressed in the edge channel, it can be considered that an increase in the source–drain current corresponds to an increase in the number of harmonic oscillators: $\Delta \tau = ne/nI_c = e/I_c$. Thus, as the source–drain current increases, the classical relation $I = 2\pi eNf$ is satisfied for a fixed generation frequency, where N is the number of generating oscillators (Fig. 2). We can see that the dependence of the conductance on the source– drain current reflects the process of their sequential actuation (Fig. 2).

In order to better understand the characteristics of the transport of current carriers in the edge channels of the ultra-narrow quantum well, which forms the basis of the silicon nanosandwich, in particular, the Josephson effects, an experiment was conducted, where SNS was used as a recorder and source of microwave radiation. To register the THz response in this case, the identification of Shapiro steps under conditions of applying voltage in the plane of the quantum well and, in particular, along and across its edge channels, inside which the quantum spin-dependent carrier transport is implemented, was used. Thus, a pair of silicon nanosandwiches was used, one of them as a source, and the other as a recorder of microwave radiation. The Shapiro step when measuring the voltage on the $U_{xx}$ (source-drain) contacts of the recorder arises as a resonance when it is irradiated with a nanosandwich acting as a source, and the resonance is recorded by measuring the voltage change on the $U_{xx}$ (source-drain) contacts of the recorder. In this case, the resonant response of the recorder is a consequence of a change in the electromagnetic field of the source: $h\nu=2eU_{xx}$. During the experiment, the longitudinal current of the source-drain recorder was set, and a voltage drop was recorded at the contacts $U_{xx}$. Then, the stabilized source-drain current is changed in time, which represents also the source of an external electromagnetic field. Under these conditions, simultaneously with the switching on of the source current, a $U_{xx}$ response arised, which seems to be analog the Shapiro step.

Thus, the frequency of electromagnetic radiation: $U_{xx}*2e/h=\nu$ appears to be determined by measuring the magnitude of the response, $U_{xx}$. In addition, based on the obtained $U_{xx}$ value, it is possible to find out the resistance grid both at the recorder and at the source, within the framework of which the Shapiro step was revealed. In particular, the experimental result indicates that the edge channel of a silicon nanosandwich, whose length 4.72 μm corresponds to a frequency of 9.3 GHz, acts as a resonator in the GHz range. It should be noted that this result is confirmed by the use of electrical detection of EPR in the 3-centimeter range in an external magnetic field on a given nanosandwich under the conditions of the source-drain current stabilization. Previously, the resonance response has been shown to be revealed under the conditions of tunnel junctions between heavily doped δ-barriers [13, 14]. It should be emphasized that the response has always been nonlinear depending on the current source [16]. Great interest is the study of the resonant response at a frequency of 2.8 THz, which is of great practical importance for use in medicine. Finally, the experimental procedure can be modeled by replacing the THz emitter with a DNA oligonucleotide deposited on the surface of the recorder (see Fig. 1c). Quantum harmonic oscillators containing single holes inside the SNS edge channels are THz-generation sources with increasing source–drain current, with the amplification by the insertion of microcavities [18, 19]. Therefore, THz-radiation spectra corresponding to the frequency range of eigenmodes of oligonucleotides deposited onto the SNS surface can be formed by varying the microcavity sizes [20-22]. The response of these eigenmodes is a change in the I–V characteristic of the edge-channel conductance [18]. Furthermore, the effect of feedback of the influence of the oligonucleotides on the SNS conductance can be enhanced, if the microcavity contains a single hole and one oligonucleotide [18]. For example, the eigenmode frequency of the oligonucleotide firstly studied using the SNS, i.e., 2.8 THz, corresponds to a microcavity size of 16.6 μm. Thus, to amplify the terahertz response of an oligonucleotide in the I–V characteristic of the SNS, the average distance between holes in the SNS edge channel should be close to this value. To this end, the density of two-dimensional holes in the SNS was optimized near a value of $3 \times 10^{13}$ m$^{-2}$ which defines the number of carriers between measuring contacts "xx" ~ 120 which is consistent with the above-mentioned average distance between them, i.e., 16.6 μm. Such optimization of the average distance between holes in the SNS edge channel according to the embedded microcavity sizes made it possible to develop efficient THz-radiation sources based on edge-channel fragments containing single holes (Fig. 3). In this case, the characteristics of these quantum harmonic oscillators are reflected not only in the detection of Rabi splitting (see Fig. 3), but also in the detection of the gigahertz modulation of THz-radiation spectra, which is controlled, as noted above, by the edge-channel length.

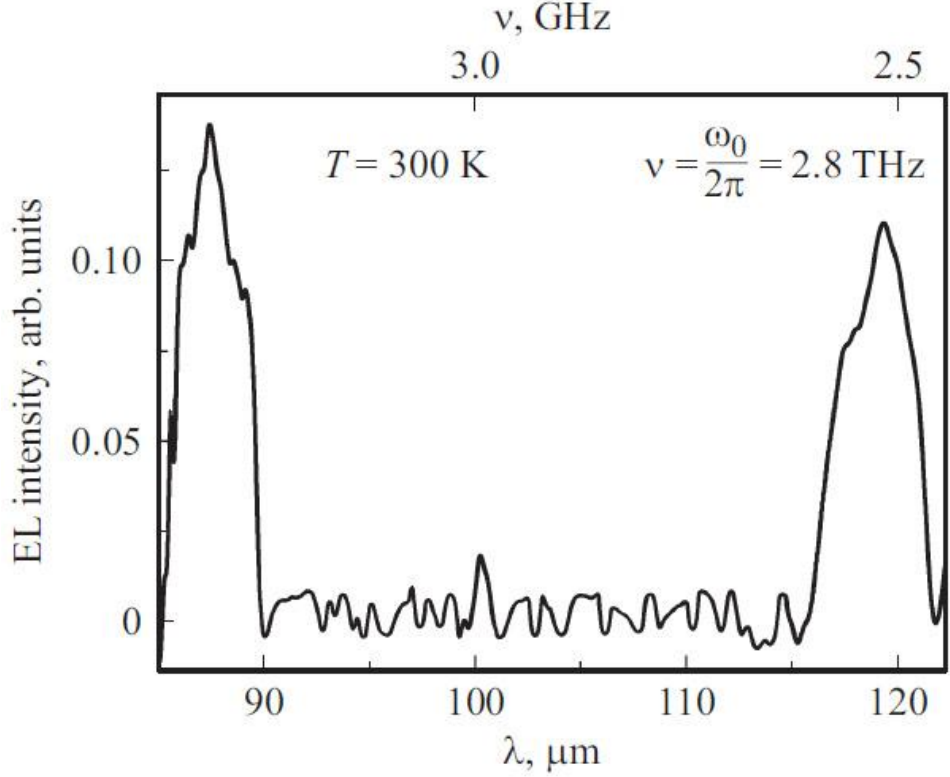

**Fig. 3.** Electroluminescence spectrum of a SNS with embedded microcavities, showing the contribution of the microcavity with a feature size of 16.6 μm and Rabi splitting under conditions of THz generation by the quantum harmonic oscillator at a frequency of 2.8 THz at room temperature.

### Materials

The *I–V* characteristics of the SNSs within Hall geometry were measured at a stabilized source–drain current (Figs. 4,5) to determine the resonant frequencies of oligonucleotides (Fig. 1a). Oligonucleotide molecules were precisely deposited onto the δ-barrier over SNS edge channels using a Proline plus Biohit micropipette (0.1–3 μL) and a container-type microfluid system (Figs. 1a, 1c, 1d) made of dimethyl silicone and placed on the SNS surface. Its volume contained 0.5 μL of solution, preventing its evaporation during device operation. Single-stranded oligonucleotide molecules were synthesized using an Applied Biosystems oligonucleotide synthesizer by the phosphoramidite method, purified by electrophoresis in a polyacrylamide gel, and extracted in a 0.3-M solution of sodium acetate. The following oligonucleotide sequence was studied: 100-mer 5'-gcgctggctgcgggcggtgagctgagctcgcccccggggagctgtggccggcgcccct gccggttccctgagcagcggacgttcatgctgggagggcggcg-3'.

The concentrations of oligonucleotide molecules were selected so that there was no more than one oligonucleotide molecule for each microcavity, and their values would be 0.22 and 0.98 μg/μL, respectively. These concentration parameters were chosen to satisfy the relation with the number of holes in the edge channels. To perform comparative analysis, SNSs without oligonucleotides were also studied, since, as shown in [18], the buffer solution does not cause significant changes reflected in the *I–V* characteristics of the SNS.

### Results and discussion

The dependences $U_{xx} - I_{ds}$ and $U_{xy} - I_{ds}$ were studied at room temperature using a dc source (Keithley 6221), two nanovoltmeters $U_{xx}$ and $U_{xy}$ (Keithley 2182A), and a grounded metal container with a sample holder.

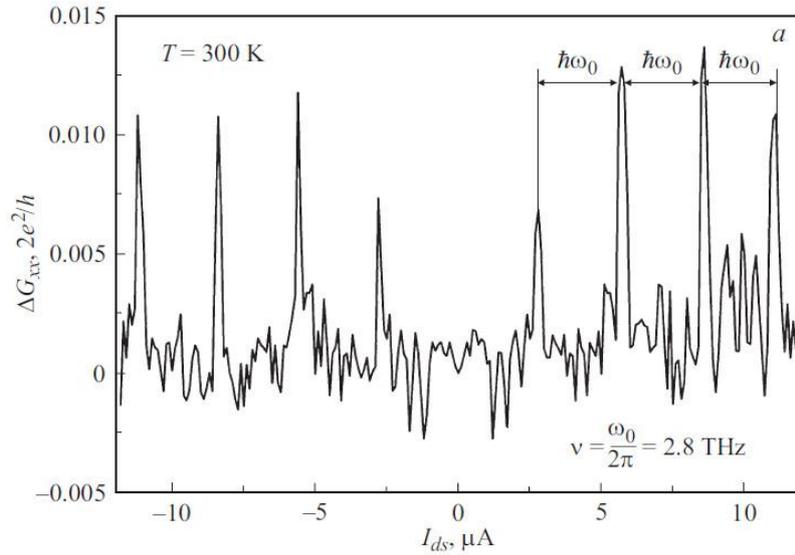

**Fig. 4.** Response of eigenmodes of single-stranded 100-mer oligonucleotides in the SNS edge channel, manifesting itself in a change in the dependence of the conductance on the source–drain driving current under conditions of THz generation.

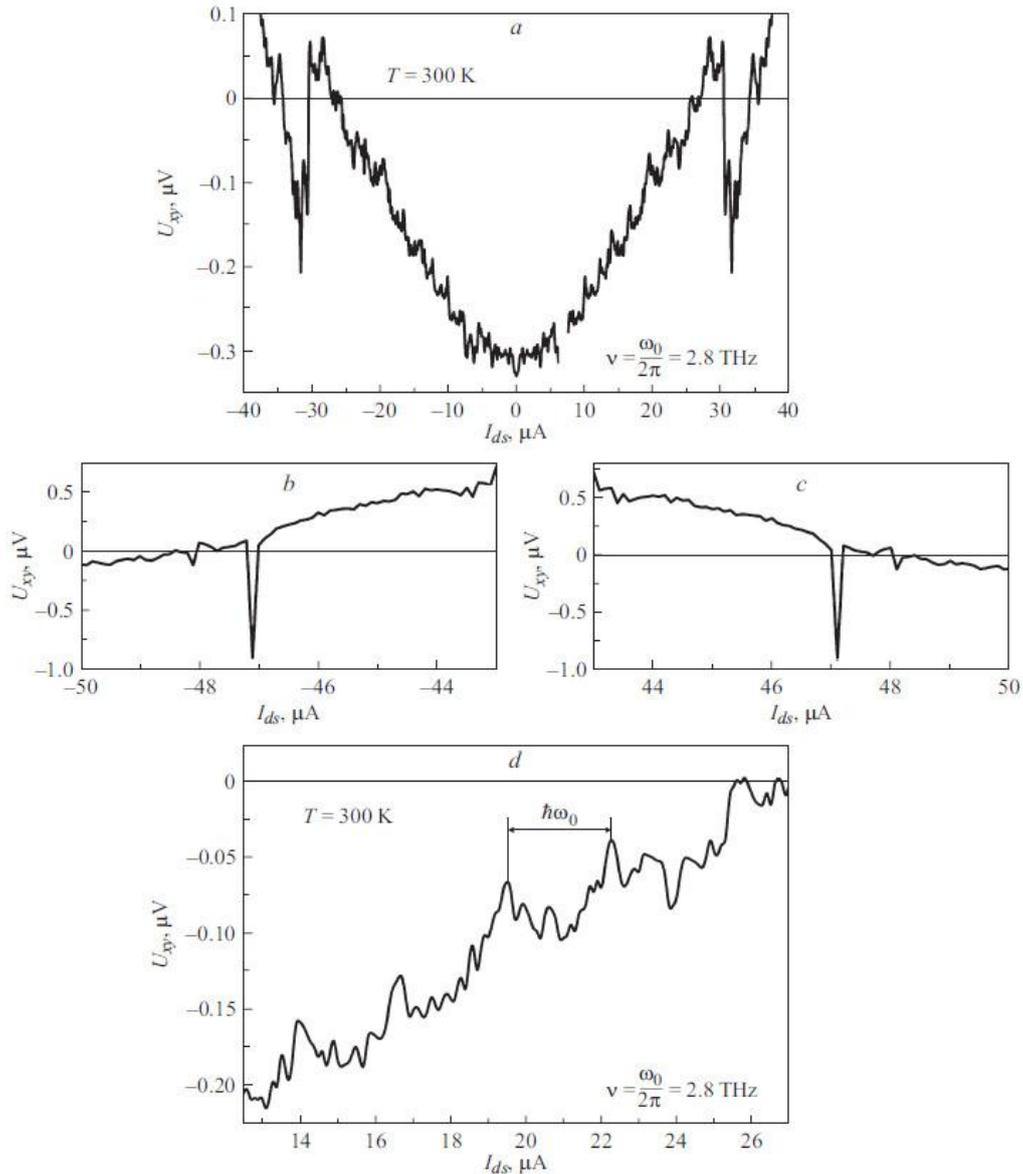

**Fig. 5.** Dependence of the transverse voltage $U_{xy}$ of SNS with deposited DNA oligonucleotide on the driving current (a-d).

The system was synchronized using a National Instruments Lab View environment. The ranges of stabilized driving current (–50–50 µA) with a step of 100 nA and (–3–3 µA) with a step of 20 nA. Each point was measured 10 times with an interval of 1 ms.

As a result of measurements of the *I–V* characteristics of the SNS, the change in the conductance was recorded in units of $G_0 = 2e^2/h$. This dependence on the source–drain current $I_{ds}$ exhibits periodic beats (Fig. 4). Having analyzed the dependences presented in Fig. 4, the frequency corresponding to the resonance response of DNA oligonucleotide the THz radiation of the microcavity containing a single hole can be determined from the beat period. It appeared that the frequency of the resonance response of the DNA oligonucleotide, 2.8 THz, determined from the beat period using the uncertainty relations is in good agreement with the data preliminarily obtained at a stabilized source–drain current of 48.1 µA [18]. Taking into account the frequency dependence of the source– drain driving current, $I = 2\pi eNf$, the resonance response of oligonucleotides at $I_{ds}$ = 48.1 µA appears in the case of in-series connected 17 microcavities whose THz radiation is responsible for the excitation of DNA oligonucleotides.

Since the frequency of the THz response of the oligonucleotide almost coincides with the resonance frequency of the microcavity in which it is placed, being deposited onto the surface of the SNS edge channel, the contribution of other harmonics to the observed conductance beats is barely manifested in the *I–V* characteristic. To be convinced of this, the *I–V* characteristics of the SNS, measured in the absence and in the presence of oligonucleotides on the SNS surface, should be compared in more detail. Therewith, besides analysis of the *I–V* characteristic of the conductance, information about the frequency of the resonance response of the oligonucleotide can be obtained by studying the dependence of the transverse voltage $U_{xy}$ on the driving current, which also exhibits periodic beats (Figs. 5a–5d), whose period is identical to the period of beats depending on the conductance (Fig. 4), corresponding to the frequency of the resonance response of the oligonucleotide $f_{res}$ = 2.8 THz.

The eigenfrequencies of quantum harmonic oscillators manifest themselves on the SNS surface in the absence of DNA oligonucleotides, which are formed by single holes within fragments of edge channels confined by chains of dipole centers with negative correlation energy. In this case, beats appearing in the dependence of the conductance on the source–drain driving current correspond to the frequencies λ ~ 2.1 THz and λ/2 ~ 4.2 THz. In this case, the maximum beat frequency corresponds to the second harmonic of THz radiation of the quantum harmonic oscillator, λ/2, which more optimally corresponds to the resonance frequency of the embedded microcavity. It should be noted that the THz-generation frequency is different without and with oligonucleotides deposited onto the surface of the SNS edge channels due to the effect of the electron–electron interaction of holes forming neighboring oscillators, which is depressed during their interaction with oligonucleotides. In the latter case, if the oligonucleotide is in the cavity, the harmonic oscillator is formed near the oligonucleotide. This results in the additional depression of the electron–electron interaction, and the microcavity frequency becomes dominant in the formation of the period of observed conductance beats and the transverse potential difference (2.8 THz). Also in some cases, the manifestation of various combinations of conductance and transverse voltage beats caused by direct THz radiation from SNS edge channels and the resonance response of oligonucleotides deposited onto their surface is possible, which probably results from their weak bond with embedded microcavities.

By analogy with experiments on the resonant response of the oligonucleotide as a reflector of the THz radiation, a bio-tissue is able to be used to generate the THz resonance response to the current-voltage characteristics of silicon nanosandwiches that appears to be applied for the express diagnostics of some diseases. It is important to note that previous studies confirm the possibility of diagnosing cancer using devises working in the THz frequency range [8, 22]. Despite all the success all experiment are conducted with huge expensive setups and most of them are limited laboratory animal experiments. Previously described SNS system showed that it is capable not only to work as a generator of THz radiation but also be a recorder at the same time. Mechanism of registration of the Shapiro steps used to describe frequency characteristics of DNA applied to the sample can also be used to identify properties of bio-tissue exposed to THz radiation from SNS [14].

For THz cancer express diagnosis research we built a spectrometer with a SNS, built in the Hall geometry, used as a THz generator. During experiments current-voltage characteristics were measured. Device was aimed to the point of neoplasm localization. To find out correlations between reflection and/or emission properties of bio-tissue repeated measurement of the points remote from initial were conducted. As was said earlier, by the analogy with the experiments with oligonucleotides, bio-tissue, in this case, performed as an emitter whereas SNS was a recorder of THz radiation. By the other words, device operates as a balance recorder. I.e., current-voltage characteristics of the device carry information about bio-tissue properties which is shown in the dependence $U_{xx}$ on the stabilized drain-source current interconnected with frequency (Fig 6). The dependence shows three different cases that correspond to different stages of oncology of the female breast. The control was carried out using ultrasound and high-resolution x-ray methods. It is clear that the signal power increases with the development of cancer. In addition to the contribution of the DNA oligonucleotide with a frequencies of adenine(3.2 THz), guanine(2.9 THz), cytosine(2.7 THz) and thymine(2.5 THz) in the process of developing oncology, other features appear that indicate deterioration of the lymphatic system. Moreover, the phase of the local THz current-voltage characteristic is largely determined by the site of radiation effect on the biological tissue. In addition, there is a characteristic change in the current-voltage characteristics in the region of one of the base frequencies (160 GHz), which reflects the operation of the lymphatic system. In particular, a change in the phase of the signal to a negative, as well as a shift of the peak to the frequency range of about 120 GHz, indicates an increasing activity of cancer. The observed analogous phase change in the IVC signal in the 3 THz region is interrelated with the patient's DNA structure, with carrier tunneling on adenine-thymine bonds. I.e., spectral characteristic represents human genome. A change in the phase of the signal on the resonant frequencies for the patient indicates an increasing activity of the tumor. It is a basis of more detailed THz diagnostics in future.

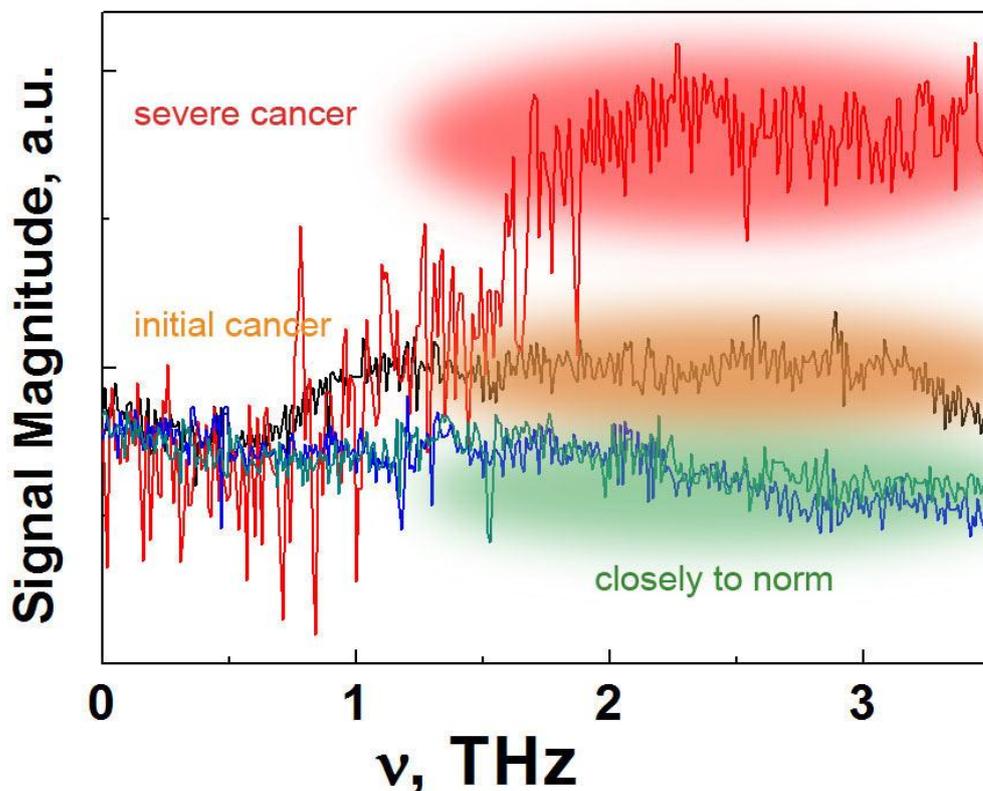

**Fig. 6.** The THz response revealed by the conductance of the silicon nanosandwich which is obtained from the balance of radiation of biological tissue and silicon nanosandwich emitter. These spectral characteristics allow us to classify various pathological conditions of the tissue.

**Summary**

It was shown that oligonucleotides can be identified by studying the change in the conductance and transverse potential difference of a SNS under conditions of their application on the SNS edge-channel region. From the value of the oscillation period of the longitudinal conductivity and the transverse potential difference ladder at Hall contacts, the magnitude of the frequency response of the DNA oligonucleotide deposited on the surface of a silicon nanosandwich is determined. Thus, the development of this technique can be used to determine various characteristics of DNA oligonucleotides, in particular, to identify their dielectric, magnetic and optical properties, that is important when the THz radiation sources are applied to for various purposes of personalized medicine, as well as for developing express diagnostics of DNA oligonucleotides.

**References**

1. C. Toumazou, et al., Nature Methods, 10, 641 (2013).
2. E.R. Mardis, Annual Review of Analytical Chemistry, 6, 287 (2013).
3. J.M. Rothberg, et al., Nature, 475, 348 (2011).
4. C. Plesa, N. van Loo, and C. Dekker, Nanoscale 7, 13605 (2015).
5. M. Wanunu, Phys. Life Rev. 9, 125 (2012).
6. M. Jain, I. T. Fiddes, K. H. Miga, H. E. Olsen, B. Paten, and M. Akeson, Nat. Methods 12, 351 (2015).
7. M. Muthukumar, C. Plesa, and C. Dekker, Phys. Today 68 (8), 40 (2015).
8. K. Humphreys, J. P. Loughran, M. Gradziel et al., Medical applications of Terahertz Imaging: a Review of Current Technology and Potential Applications in Biomedical Engineering, *Proc. of 26th Annual Int. Conf. of the Engineering in Medicine and Biology Society,* 2004, p. 1302.
9. N. T. Bagraev A. Bouravleuv, W. Gehlhoff, L. Klyachkin, A. Malyarenko, and S. Rykov, Def. Dif. Forum 194, 673 (2001).
10. N. T. Bagraev, N. G. Galkin, W. Gehlhoff, L. E. Klyachkin, and A. M. Malyarenko, J. Phys.: Condens. Matter 20, 164202 (2008).
11. N. T. Bagraev, A. D. Bouravlev, L. E. Klyachkin, A. M. Malyarenko, V. Gehlhoff, V. K. Ivanov, and I. A. Shelykh, Semiconductors 36, 439 (2002).
12. N. T. Bagraev, A. D. Bouravlev, L. E. Klyachkin, A. M. Malyarenko, V. Gehlhoff, Yu. I. Romanov, and S. A. Rykov, Semiconductors 39, 685 (2005).
13. N. T. Bagraev, W. Gehlhoff, L. E. Klyachkin, A. A. Kudryavtsev, A. M. Malyarenko, G. A. Oganesyan, D. S. Poloskin, and V. V. Romanov, Physica C 219, 437 (2006).
14. N. T. Bagraev, V. A. Mashkov, E. Yu. Danilovsky, W. Gehlhoff, D. S. Gets, L. E. Klyachkin, A. A. Kudryavtsev, R. V. Kuzmin, A. M. Malyarenko, and V. V. Romanov, Appl. Magn. Reson. 39, 113 (2010).
15. N. T. Bagraev, E. Yu. Danilovsky, L. E. Klyachkin, A. M. Malyarenko, and V. A. Mashkov, Semiconductors 46, 75 (2012).
16. N. T. Bagraev, L. E. Klyachkin, A. A. Kudryavtsev, A. M. Malyarenko, and V. V. Romanov, in Superconductor, Ed. by A. Luiz (SCIYO, Rijeka, 2010), p. 69.
17. N. T. Bagraev, V. Yu. Grigoryev, L. E. Klyachkin, A. M. Malyarenko, V. A. Mashkov, and V. V. Romanov, Fiz. Tekh. Poluprovodn. 50 (8) (2016, in press).
18. A. L. Chernev, N. T. Bagraev, L. E. Klyachkin, A. K. Emel'yanov, and M. V. Dubina, Semiconductors 49, 944 (2015).
19. N. T. Bagraev, E. Yu. Danilovskii, and D. S. Gets, J. Phys.: Conf. Ser. 486, 012017 (2014).
20. G. J. Thomas, Jr., Ann. Rev. Biophys. Biomol. Struct. 28, 1 (1999).
21. A. Barhoumi, D. Zhang, F. Tam, and N. G. Halas, J. Am. Chem. Soc. 130, 5523 (2008).
22. B. M. Fischer, M. Walther, and P. Jepsen, Phys. Med. Biol. 47, 3807 (2002).